%% file: cryoMBIR.tex
\documentclass[12pt,onecolumn,draftcls]{IEEEtran}                                                                               
\include{package}

\include{defs}

\title{Model-based Reconstruction for Single Particle Cryo-Electron Microscopy}

\author{S.~V.~Venkatakrishnan$^{\star}$,
  \thanks {$^{\star}$ Multi-modal Sensor Analytics Group, Oak Ridge National Laboratory, Oak Ridge, TN 37831, USA}
  Puneet Juneja $^{\dagger}$
  \thanks{$^{\dagger}$ Iowa State University, Ames, IA 50011, USA},
  Hugh O'Neill $^{o}$
  \thanks{$^{o}$ Neutron Scattering Division, Oak Ridge National Laboratory, Oak Ridge, TN 37831, USA}
 \thanks{\scriptsize{This abstract has been authored by UT-Battelle, LLC., under Contract No. DE-AC05-00OR22725 with the U.S. Department of Energy.
The United States Government and the publisher, by accepting the article for publication, acknowledges that the United States Government retains a non-exclusive, paid-up, irrevocable, world-wide license to publish or reproduce the published form of this manuscript, or allow others to do so, for United States Government purposes.
DOE will provide public access to these results of federally sponsored research in accordance with the DOE Public Access Plan (http://energy.gov/downloads/doe-public-access-plan).
}}
}

\begin{document}

\maketitle

\input{front}

\input{cryoMBIRBody}

\bibliographystyle{IEEEbib}
\footnotesize
\bibliography{cryoMBIR}

%%%%%%%%%%%%%%%%%%%%%%%%%%%%%%%%%%%%%%%%%%%%%%%%%%%%%%%%%%%%%  

\end{document}

%% file: package.tex
\usepackage{array}
\usepackage{mdwmath}
\usepackage{mdwtab}
\usepackage{fixltx2e}
\usepackage{url}
\usepackage{color}
\usepackage[]{graphicx}
\usepackage{epsfig}
\usepackage{bbm}
\usepackage{enumerate}
\usepackage{amsfonts}
\usepackage{amsmath}
\usepackage{amssymb}
\usepackage{cite}
\usepackage{algorithm}
\usepackage{algpseudocode}
\usepackage{eqparbox}
\usepackage{comment}
\usepackage{breqn}
\usepackage{dblfloatfix}

%% file: defs.tex
\providecommand{\norm}[1]{\lVert#1\rVert}

\DeclareMathOperator*{\argmin}{argmin}

%% file: front.tex
\begin{abstract}
  Single particle cryo-electron microscopy is a vital tool for 3D characterization of protein structures. 
  A typical workflow involves acquiring projection images of a collection of randomly oriented particles, picking and classifying individual particle projections by orientation, and finally using the individual particle projections to reconstruct a 3D map of the electron density profile. 
  The reconstruction is challenging because of the low signal-to-noise ratio of the data, the unknown orientation of the particles, and the sparsity of data especially when dealing with flexible proteins where there may not be sufficient data corresponding to each class to obtain an accurate reconstruction using standard algorithms. 
  In this paper we present a model-based image reconstruction technique that uses a regularized cost function to reconstruct the 3D density map by assuming known orientations for the particles. 
  Our method casts the reconstruction as minimizing a cost function involving a novel forward model term that accounts for the contrast transfer function of the microscope, the orientation of the particles and the center of rotation offsets. 
  We combine the forward model term with a regularizer that enforces desirable properties in the volume to be reconstructed. 
  Using simulated data, we demonstrate how our method can significantly improve upon the typically used approach.
\end{abstract}

%  Neutrons are useful for imaging because they can penetrate thick specimens and provide
%  a complimentary contrast to X-rays.

%% file: cryoMBIRBody.tex
\section{Introduction}
\begin{figure*}[!htbp]
\begin{center}
    \includegraphics[scale=0.5,trim=0.2cm 13cm 0cm 0cm,clip]{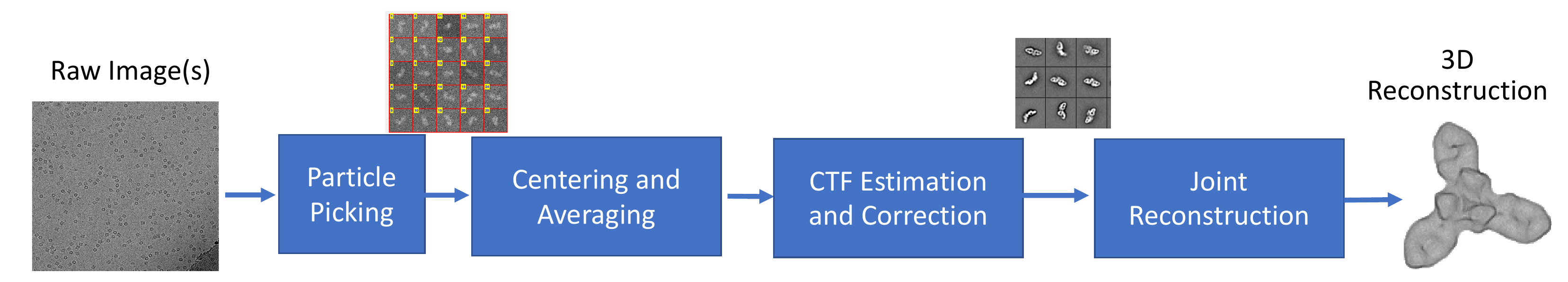}
\end{center}
\caption{\label{fig:pipeline} Simplified illustration of the processing pipeline for SPR. 
The central challenge is one of reconstructing the 3D structure of an object from extremely noisy data that corresponds to the projection of the original object at certain unknown orientations.
}
\end{figure*}
%What is the problem ? 
Single particle reconstruction (SPR) for cryogenic electron microscopy (cryo-EM) involves determining the 3D structure of macro-molecules from projection images of randomly oriented replicates of these particles which are flash frozen in vitrified ice and imaged using an EM \cite{sigworth2016principles}.
A typical reconstruction workflow (see Fig.~\ref{fig:pipeline}) involves picking the particles from an image containing the projection of a large number of these particles, centering the picked particles, clustering and averaging similar particles to boost the SNR corresponding to a certain orientation, deconvolving these resulting images which are impacted by the contrast transfer function of the microscope (CTF correction), followed by an iterative scheme that jointly estimates the orientation and reconstructs the 3D volume \cite{sigworth2016principles}.
Due to dose limitations, the data is extremely noisy making it challenging to obtain high-quality single particle reconstructions. 

The tomographic reconstruction (for a fixed set of orientations)
is often done using a direct/iterative Fourier method \cite{penczek2004gridding,wang2013fourier,abrishami2015fast}, because it is fast and hence appealing to use in an iterative refinement procedure.
However, such methods can result in severe artifacts in the presence of noise and the absence of
a uniform sampling of  projection orientations because of the preferential
orientation of particles \cite{baldwin2019non}.
Furthermore, there is increasing interest to study flexible protein structures
\cite{villarreal2014cryoem}, which consist of particles from different conformations in the data
resulting in fewer overall orientations and more noise when particles are averaged.
Finally, even if the standard methods are used to reconstruct a particle,
the reconstructions can be significantly improved with a final
reconstruction step that uses the estimated orientation with the \textit{raw
noisy measurements} to obtain a reconstruction using a more advanced
method than the direct Fourier techniques as has been demonstrated in a wide variety of electron tomography applications \cite{VenkatHAADF13,VenkatBF15,yan2019mbir}. 

%Iterative methods
While direct or iterative Fourier methods are predominantly used for SPR \cite{penczek2004gridding,wang2013fourier,abrishami2015fast}, a few
model-based/regularized iterative methods have been proposed to improve the reconstruction step.
These methods solve the reconstruction by minimizing a cost function that
balances a data-fidelity term based on a forward model and a regularization term based on some assumptions about the underlying object itself.
Liu et al. \cite{li2011single} presented a reconstruction algorithm (for known particle orientations) by using a quadratic data-fitting term along with a total-variation regularizer applied to coefficients in spline-basis.
However, this work does not take into account the contrast transfer function of the microscope and the offset of the particles with respect to the center of the projections in the forward model. 
Kuckukelbir et al. \cite{kucukelbir2012bayesian} used an adaptive wavelet basis along with a $l_{1}$ regularizer on the coefficients to illustrate how the reconstruction can be improved compared to traditional methods.
Pan et al. \cite{pan2018fast} solve the reconstruction using a total-variation prior, while Donati et al. \cite{donati2018fast,donati2019inner} formulate a regularized cost function using a spline basis that allows for fast multi-scale reconstruction.
Zehni et al. \cite{zehni2020joint} developed a regularized iterative reconstruction technique that also takes into account the joint-estimation of the angles in addition to the 3D reconstruction by using a radial-basis function to parameterize the volume and a total-variation regularizer for the coefficients. 
In summary, there have been a few efforts at leveraging the success of model-based/regularized iterative techniques to improve single particle reconstructions. 

%Proposed approach
In this paper, we present a model-based image reconstruction (MBIR) approach based on minimizing a  regularized cost function \cite{VenkatBF15} for solving the single particle cryo-EM problem \textit{for a known set of particle orientations}.
This method can be used within a refinement loop or applied as a final step to the raw measurements in order to obtain a high quality reconstruction from noisy, and limited orientation data sets. 
In contrast to the methods in \cite{li2011single,pan2018fast,donati2018fast} that rely on a spline basis, we use a simple voxel basis with projectors implemented to work with graphic processing units (GPU). 
Our forward projectors includes a model for center-of-rotation offsets and the contrast transfer function of the microscope, thereby avoiding the need to pre-process the data which can result in a loss of resolution.
%This approach is fundamentally different from the traditional approach of centering and CTF correction followed by reconstruction since we model the CTF and object offset into the image reconstruction itself.
Furthermore, the proposed forward model also allows for modeling of non Gaussian noise in the data; which is more accurate for the extremely low SNR count data that is encountered in cryo-EM detectors.
Furthermore, instead of restricting ourselves to  a $l_{1}$ or TV regularizer \cite{li2011single,kucukelbir2012bayesian,pan2018fast,donati2018fast,zehni2020joint}, we use a generalized Markov random field \cite{JBSaBoHsMultiSlice} based regularizer allowing for a broader range of solutions.  
We demonstrate the utility of our algorithm on realistic simulated data sets and highlight the utility of the method compared to the pre-process and reconstruct approach of Fig.~\ref{fig:pipeline}.

\section{Model-based Image Reconstruction}
\label{sec:mbir}
\begin{figure}[!htbp]
\begin{center}
    \includegraphics[scale=0.65,trim=0.2cm 0.5cm 16cm 1.5cm,clip]{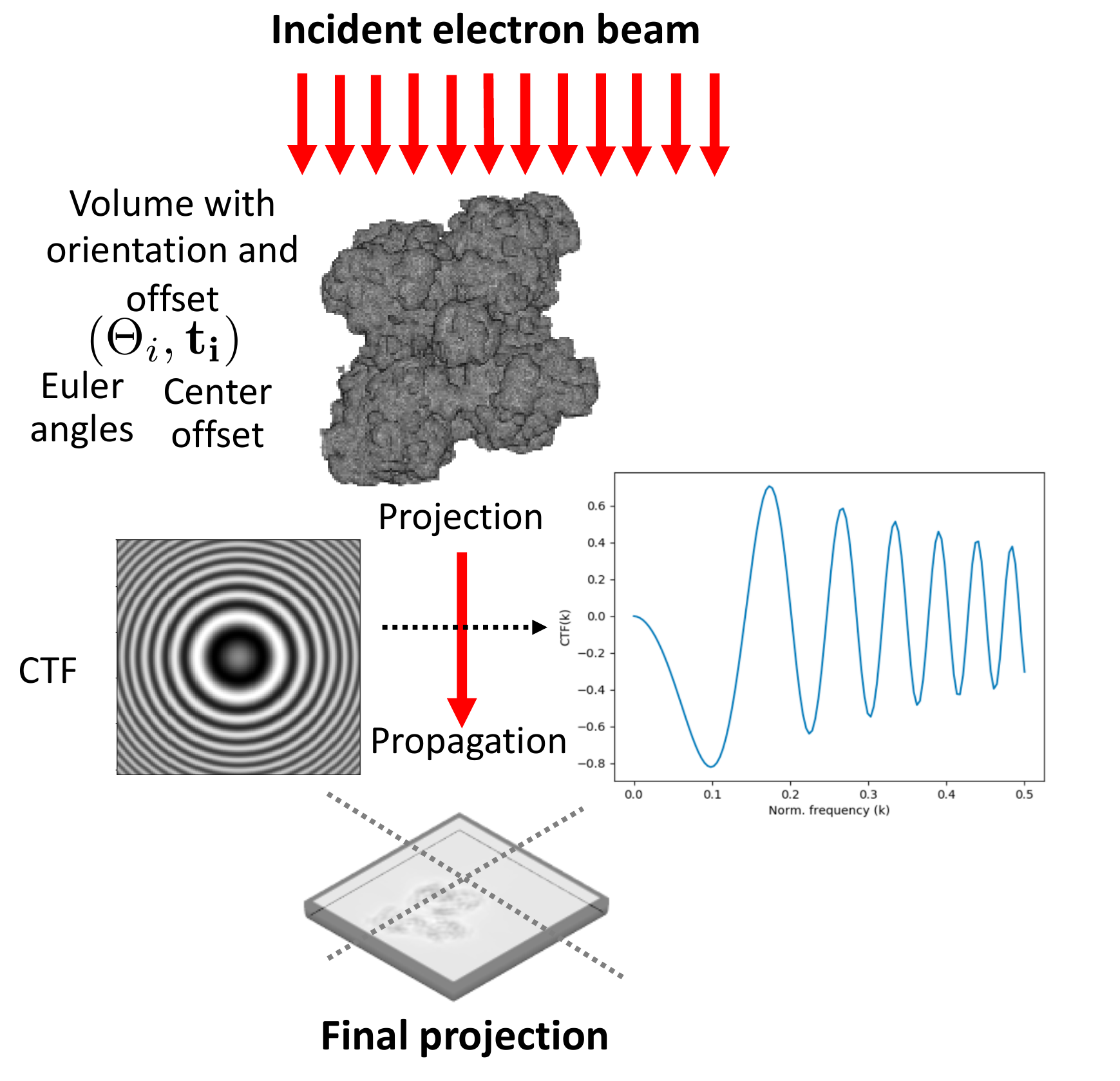}
\end{center}
\caption{\label{fig:forward_model} Illustration of the forward model used for the cryo-EM MBIR method. 
It involves a 3D projection at a fixed orientation (with appropriate offset for the center of rotation) followed by a propagation operator that depends on the contrast transfer function (CTF) of the system.
The figure shows an example of the magnitude of the Fourier transform of a typical CTF, illustrating that the CTF typically \textit{zeros out} several frequency components of the projection data. 
While this can pose challenges for typical pre-process and reconstruct approaches, we use this model in order to perform the reconstruction. 
}
\end{figure}

In order to reconstruct the density in 3D,
we use the MBIR \cite{Bo641Text} framework.
The reconstruction is formulated as a minimization problem, 
\begin{eqnarray}
\label{eq:MBIRCost}
\hat{f} \leftarrow \argmin_{f} \left\{ l(g;f) + s(f)\right\}
\end{eqnarray}
where $g$ is the vector of projection measurements, $f$ is the 
vector containing all the voxels,
$l(;)$ is a data fidelity enforcing function 
and $s(.)$ is a function that enforces regularity in $f$. 
%Forward model
To formulate the data fidelity term, we use the physics-based model (see Fig.~\ref{fig:forward_model}) where each measured image is modeled as the projection of the unknown object at a specific orientation and offset, followed by a propagation effect due to the contrast transfer function of the microscope.
Since the measurements are typically corrupted by noise that depends on the acquisition dose, we propose a quadratic data-fidelity term of the form 
\begin{eqnarray}
  l(g;f) = \frac{1}{2}\norm{g-HAf}_{W}^{2}
  \label{eq:LLD}
\end{eqnarray}
where $H$ is a matrix modeling the contrast transfer function (CTF) of the imaging system as a linear shift invariant filter, $A$ is a forward projection matrix that accounts for the
3D orientation ($\Theta_{i}$) of the particles and offsets ($\text{t}_{i}$) of the projections from the center of the projection images, and $W$ is a diagonal matrix with entries set to be the inverse variance of the noise in $g$ (``dose weighting'').
The $W$ matrix can also be used to mask regions of the measurements that are corrupted due to various other reasons (like overlapping particles), providing an additional flexibility to the reconstruction.  
%This choice of data-fidelity term corresponds to a forward model of the form as illustrated in Fig.~\ref{fig:forward_model}.
%The entries of $W$ are set such that $W_{ii}=\lambda_i$, where $\lambda_i$ is the raw un-normalized measurement \cite{SaBo92}.
Notice, that in contrast to existing approaches which apply ``centering'' and ``CTF correction'' to the data, our approach models these into the reconstruction itself.
Furthermore, if the data sets contains measurements made at multiple defocus
values corresponding to different CTFs, this can be simply incorporated in the model described above. 
We design $A$ to model the cryo-EM geometry by using the ASTRA tool-box \cite{AstraGPU11,AstraUltramic15} that can utilize multiple GPUs \cite{vanAarleASTRA16,BleichrodtAstra16} to accelerate the application of this matrix.
We note that despite the projection ($A$) and back-projection operators ($A^{T}$)
not being perfectly matched in ASTRA, we did not observe any specific problems with convergence of the overall algorithm.  
The CTF is assumed to the radially symmetric and is modeled as  
\begin{eqnarray}
  h(k) = \exp\{-\alpha k\} \sin \left( -\pi \Delta z \lambda k^{2} + \frac{\pi}{2}\text{C}_{s} \lambda^{3} k^4 \right)
  \label{eq:ctf}
\end{eqnarray}
where $k$ is the radial frequency component, $\alpha$ is an attenuation coefficient,
$\Delta z$ is the defocus, $\lambda$ is the electron wavelength, and $\text{C}_{s}$ is the
spherical aberration.

For $s(f)$, we choose the negative log of
q-generalized Markov-random field (qGGMRF)
probability density function
\cite{JBSaBoHsMultiSlice}. It is given by
\begin{eqnarray}
\label{eq:Prior}
s(f)&=& \sum\limits _{\{j,k\}\in \mathcal{N}}w_{jk}\rho(f_j-f_k) \nonumber \\
\rho(f_j-f_k )&=&\frac{\left|\frac{f_{j}-f_{k}}{\sigma_f}\right|^{2}}{c + 
\left|\frac{f_{j}-f_{k}}{\sigma_f}\right|^{2-p}} \nonumber
\end{eqnarray}
$\mathcal{N}$ is the set 
of pairs of neighboring voxels (e.g. a 26 point neighborhood),
$1\leq p \leq 2$, $c$ and $\sigma_f$ are
qGGMRF parameters. The weights $w_{jk}$ are 
inversely proportional to the distance 
between voxels $j$ and $k$, normalized to $1$. 
This model provides a greater degree of flexibility in the
quality of reconstructions compared to an algorithm specifically
designed for a total-variation regularizer
that may force the reconstructions to appear ``waxy'' \cite{Bo641Text}.
In particular, when $p=1$ we get a behavior similar to
a total-variation model and when $p=2$ the regularizer is a quadratic function
allowing for smoother reconstructions.

Combining the data fidelity model \eqref{eq:LLD} 
with the image model \eqref{eq:Prior} the MBIR cost function is 
\begin{dmath}
\label{eq:OriginalCost}
c(f)=\frac{1}{2}  \norm{g-HAf}_{W}^{2} + s(f) 
\end{dmath}
Thus, the reconstruction is obtained by
\begin{eqnarray*}
\hat{f} \leftarrow \argmin_{f} c(f)
\end{eqnarray*}

We use the optimized gradient method (OGM) \cite{KimOGM15}
to find a minimum of the cost function.
The algorithm involves a standard gradient computation
combined with a step-size determined using Nesterov's method.
Specifically, for each iteration $k$, 
\begin{eqnarray}
  h^{(k+1)}  \leftarrow  f^{(k)} - \frac{1}{L} \nabla c(f^{(k)}) \\
  t^{(k+1)}  \leftarrow  \frac{1+\sqrt{1+4(t^{(k)})^2}}{2} \\
  f^{(k+1)}  \leftarrow  h^{(k+1)} + \frac{t^{(k)}-1}{t^{(k+1)}}(h^{(k+1)}-h^{(k)}) \\   + \frac{t^{(k)}}{t^{(k+1)}}(h^{(k+1)}-f^{(k)})
  \label{eq:ogm_update}
\end{eqnarray}
where $t^{(0)}=1$, $L$ is the Lipschitz constant of the gradient of $c(.)$, $h^{(0)}=f^{(0)}$ is an initial estimate for the reconstruction.
The gradient of the cost-function $c(.)$ is given by
\begin{eqnarray}
  \nabla c(f) = -H^{T}A^{T}W(y-HAf) + \nabla s(f).
\end{eqnarray}
We use the ASTRA tool-box \cite{AstraGPU11,AstraUltramic15}
to implement GPU accelerated forward and back-projection operators.
For the CTF ($H$) we assume circular boundary conditions and use the FFT to accelerate the computation.

\section{Results}
\label{sec:results}

\begin{figure}[!ht]
\begin{center}
    \includegraphics[scale=0.65,trim=4cm 4cm 1.5cm 0.1cm,clip]{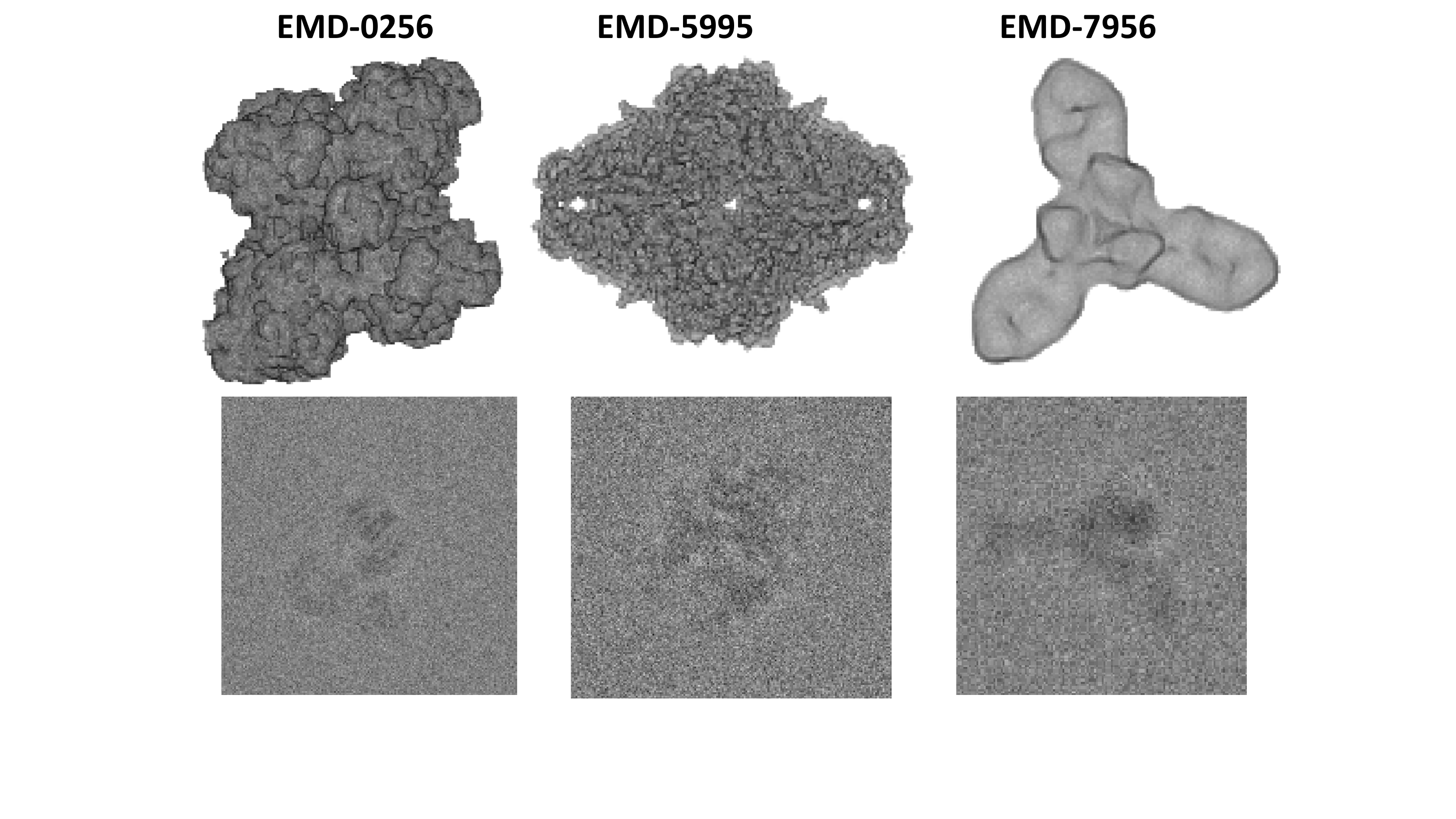}
\end{center}
\caption{\label{fig:reference_maps} 3D rendering of reference structures from the EM data bank used (top row) for generating the simulated data sets along with an example projection data (bottom row) at a peak signal-to-noise ratio of $6.02$ dB. }
\end{figure}

\begin{figure*}[!h]
 \begin{center}
  \begin{tabular}{c}
    Ground-truth \hspace{1.5in} P+R \hspace{1.5in}  MBIR \\
    \includegraphics[scale=1.4,trim={2.2cm 4.5cm 1.8cm 4.7cm},clip]{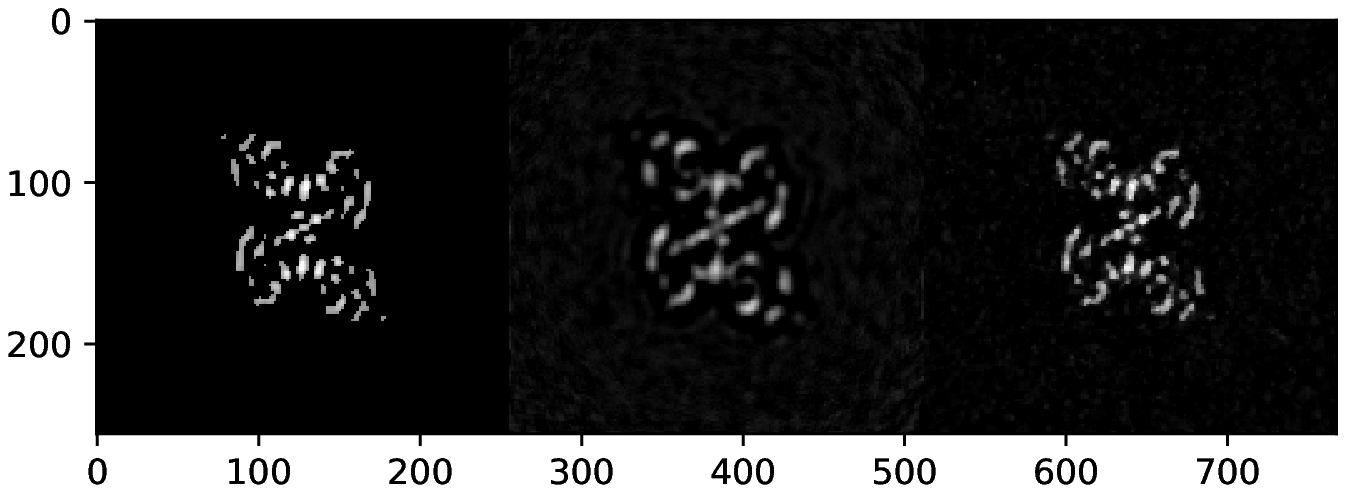} \\
    \includegraphics[scale=1.4,trim={2.2cm 4.5cm 1.8cm 4.7cm},clip]{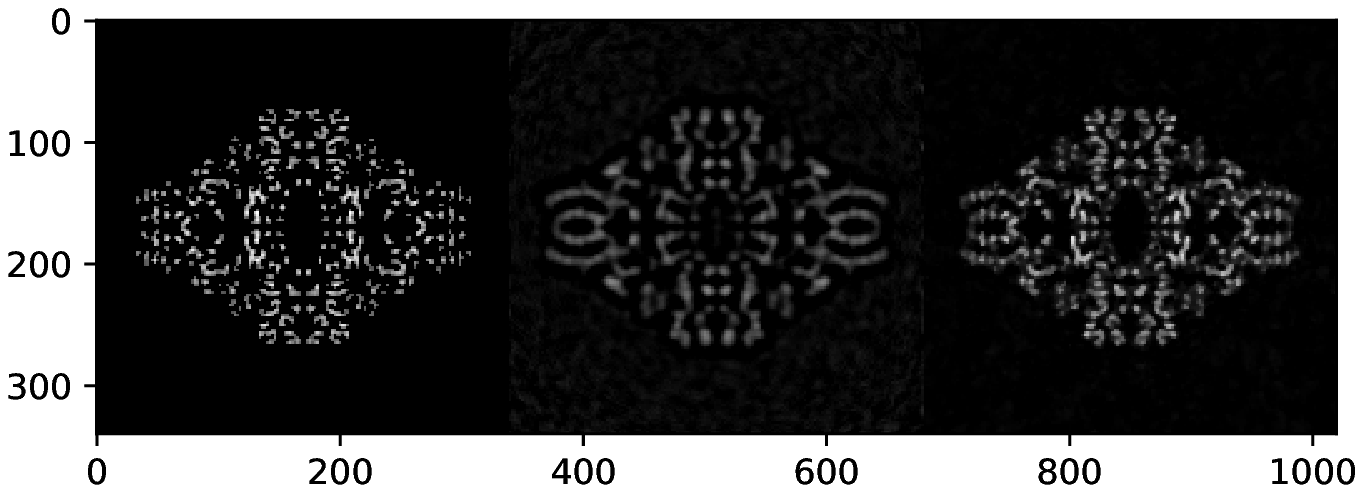} \\
    \includegraphics[scale=1.4,trim={2.2cm 4.5cm 1.8cm 4.7cm},clip]{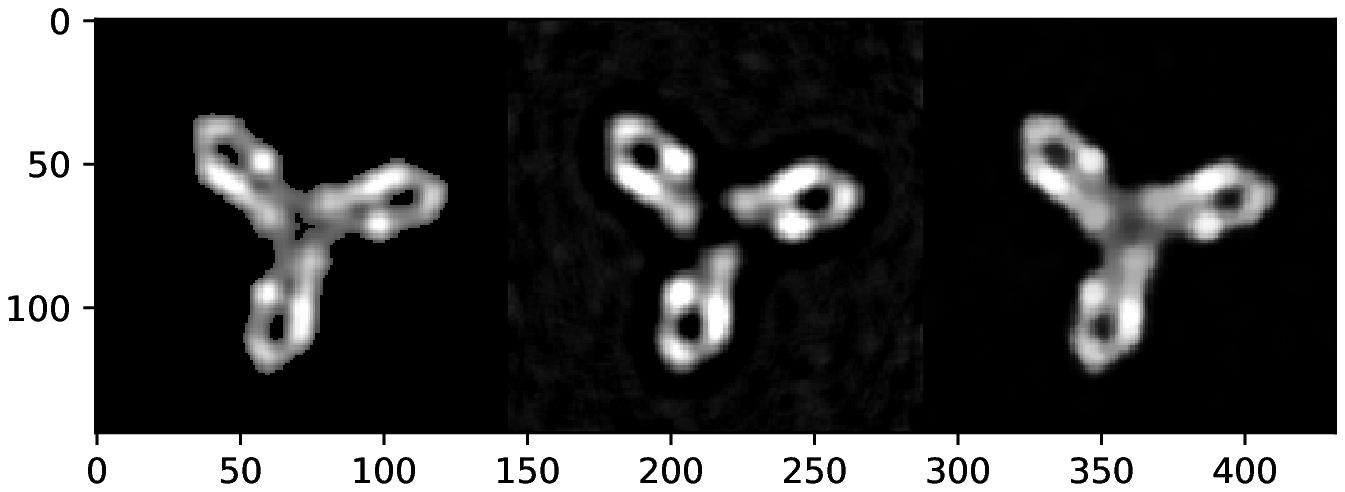}
  \end{tabular}
 \end{center}
 \caption{\label{fig:recon_sim_sparse1} A single cross section from the 3D reconstructions using different algorithms for data sets corresponding to a PSNR of 6.02 dB (Top row: EMD-0256, Middle: EMD-5995, Bottom: EMD-7956). 
 Notice that despite a very sparse data set, the proposed MBIR method can significantly improve upon the pre-process and reconstruct (P+R) approach where the reconstruction is done using a conventional algorithm.}
\end{figure*}
\begin{table}[!h]
  \caption{Comparison of normalized root mean square error (as percentage of max. density) for different reconstruction schemes and data sets identified by their EM data bank (EMD) identity number and simulated noise levels.
  %We note that the proposed MBIR approach consistently outperforms the pre-process and reconstruct (P+R) approach based on a ML technique.
  }
\label{tab:nrmse2}
\begin{center}
    All projections \\
  \begin{tabular}{| c | c |  c | c |}
    \cline{2-4}
    \hline
    Inp. PSNR  & 6.02 dB  & 2.40 dB  & 0 dB  \\
    EMD \#    &  P+R$|$MBIR & P+R$|$MBIR & P+R$|$MBIR \\
    \hline
%    \hline
    0256  & $6.65|\textbf{3.76}$  & $6.76|\textbf{4.08}$ & $6.88|\textbf{4.47}$ \\
    \hline
    5995 & $6.44|\textbf{4.04}$  & $6.55|\textbf{4.50}$ & $6.88|\textbf{4.47}$ \\
    \hline
    7956 & $4.65|\textbf{1.41} $ & $4.82|\textbf{1.55}$ & $5.03|\textbf{1.79}$ \\
    \hline
  \end{tabular}
\end{center}
\begin{center}
   50\% of projections \\
  \begin{tabular}{| c | c |  c | c |}
    \cline{2-4}
    \hline
    Inp. PSNR  & 6.02 dB  & 2.40 dB  & 0 dB  \\
    EMD \#    &  P+R$|$MBIR & P+R$|$MBIR & P+R$|$MBIR \\
    \hline
%    \hline
    0256  & $6.76|\textbf{4.08}$  & $6.94|\textbf{4.88}$ & $7.12|\textbf{5.13}$ \\
    \hline
    5995 & $6.55|\textbf{4.70}$  & $6.94|\textbf{4.88}$ & $7.12|\textbf{5.13}$ \\
    \hline
    7956 & $4.81|\textbf{1.63} $ & $5.11|\textbf{1.73}$ & $5.43|\textbf{1.90}$ \\
    \hline
  \end{tabular}
\end{center}
\begin{center}
   25\% of projections \\
  \begin{tabular}{| c | c |  c | c |}
    \cline{2-4}
    \hline
    Inp. PSNR  & 6.02 dB  & 2.40 dB  & 0 dB  \\
    EMD \#    &  P+R$|$MBIR & P+R$|$MBIR & P+R$|$MBIR \\
    \hline
%    \hline
    0256  & $6.94|\textbf{4.71}$  & $7.22|\textbf{5.49}$ & $7.48|\textbf{5.77}$ \\
    \hline
    5995 & $6.69|\textbf{5.21}$  & $6.85|\textbf{5.85}$ & $6.98|\textbf{6.17}$ \\
    \hline
    7956 & $5.11|\textbf{1.88} $ & $5.56|\textbf{2.17}$ & $5.98|\textbf{2.32}$ \\
    \hline
  \end{tabular}
\end{center}
\end{table}

In order to evaluate our algorithm, we used three structures from the EM data bank (EMD) \cite{lawson2015emdatabank} numbered 0256 \cite{emd256}, 5995 \cite{emd5995} and 7956 \cite{emd7956} to generate realistic simulated cryo-EM data sets at different noise levels and a fixed sparse number of orientations (see Fig.\ref{fig:reference_maps}).
In each case we applied the threshold recommended in the EMD, normalized the values by a constant and then simulated the projection measurements.
The volume obtained by applying the threshold and scaling serves as the ground-truth in our experiments.  
The CTF parameters (equation~\eqref{eq:ctf}) were set to $\alpha=1.0$, $\Delta z \lambda = 100$ and $\text{C}_{s} \lambda^{3}=10$.
The orientation parameters $\Theta$ were chosen so that each of the Euler angles were uniformly distributed in the $[0,2\pi]$ range leading to a preferential orientation of particles.
The offset parameters $t$ were chosen to be randomly distributed in a range of $[0,.05*p_w]$, where $p_w$ is the projected width of the simulated data in units of pixels.  
We simulated three different noise levels corresponding to a peak signal to noise ratio of $0$ dB, $2.4$dB and $6.02$ dB. 
The number of simulated projection was set to $2$ times the side length of each projection image (so if the size was $100 \times 100$, we simulated $200$ particles). 
We compared the proposed algorithm to an implementation of an pre-process+reconstruct (P+R) approach where we applied a Gaussian low-pass filter to the simulated data, followed by a phase-flipping technique \cite{singer2018mathematics} to correct for the effects of the CTF and finally reconstructing the volume using a standard least-squares type fitting technique which is a superior technique to the direct Fourier inversion techniques typically used.
In each case we adjusted the algorithm parameters to determine the values that resulted in the lowest root mean squared error (RMSE).

Fig.~\ref{fig:recon_sim_sparse1} shows the results from a single cross section
of the different reconstructions on the simulated data-sets at a noise level of 6.02 dB. 
Notice that the MBIR method can significantly improve the qualitative performance of the reconstructions compared to the P+R approach. 
We observe similar trends for the higher noise cases, but with an expected degradation of performance for all approaches. 
In order to quantify the performance of the proposed approach we present the normalized root mean squared (NRMSE) error for each of the cases (see Table.~\ref{tab:nrmse2}) illustrating the significant improvements of the MBIR method compared to the P+R approach.
We also perform the reconstructions by further sub-sampling the data set by selecting 50\% and 25\% of the original projection data and observe that the MBIR approach continues to have a lower NRMSE compared to the P+R approach (see Table.~\ref{tab:nrmse2}), highlighting that the presented approach can be very useful for cases when we have only a small number of particles to reconstruct from.

\section{Conclusion}
\label{sec:concl}
In this paper, we presented a new model-based algorithm for single particle cryo-EM reconstruction.
In contrast to existing techniques, our method casts the the reconstruction as minimizing a cost function that balances a data-fidelity term and a regularizer.
We introduced a new data-fidelity term that models the contrast-transfer function, the shift in center of rotation, the 3D tomographic projection geometry, and the noise in the data in order to accurately model the cryo-EM measurement.
Combining this with a standard Markov-random field based regularizer, we then developed an optimization algorithm based on first-order methods to find a minimum of the formulated cost function.
Using experiments from realistic simulated data sets, we demonstrated that our algorithm can dramatically improve reconstruction quality compared to traditional pre-process and reconstruct approach.

\section{Acknowledgement}

S.V. Venkatakrishnan and Hugh O'Neill were supported by Oak Ridge National Laboratory via the LDRD program.